\documentclass[10pt]{emulateapj}
\usepackage{amsmath}
\usepackage{graphicx}

\begin{document}

\title{GRB060218: A Relativistic Supernova Shock Breakout}

\author{E. Waxman\altaffilmark{1}, P. M\'esz\'aros\altaffilmark{2} \& S. Campana\altaffilmark{3}}

\altaffiltext{1}{Physics Faculty, Weizmann Institute of Science, Rehovot 76100, Israel}
\altaffiltext{2}{Dpt. of Astronomy and Astrophysics and Dpt. of Physics, 
Pennsylvania State University, University Park, PA 16802, USA}
\altaffiltext{3}{INAF-Osservatorio Astronomico di Brera, Merate (Lc), Italy}

\begin{abstract}

We show that the prompt and afterglow X-ray emission of GRB060218, as well as its early 
($t\lesssim 1$~d) optical-UV emission, can be explained by a model in which a 
radiation- mediated shock propagates through a compact progenitor star into a dense wind. 
The prompt thermal X-ray emission is produced in this model as the mildly relativistic 
shock, $\beta\approx0.85$ carrying few~$\times10^{49}$~erg,
reaches the wind (Thomson) photosphere, where the post-shock 
thermal radiation is released and the shock becomes collisionless. Adopting this 
interpretation of the thermal X-ray emission, a subsequent X-ray afterglow is predicted, 
due to synchrotron emission and inverse-Compton scattering of SN UV photons by electrons 
accelerated in the collisionless shock. Early optical-UV emission is also predicted, 
due to the cooling of the outer $\delta M\sim10^{-3}M_\odot$ envelope of the star, 
which was heated to high temperature during shock passage. The observed X-ray afterglow 
and the early optical-UV emission are both consistent with those expected in this model. 
Detailed analysis of the early optical-UV emission may provide detailed
constraints on the density distribution near the stellar surface. 

\end{abstract}

\keywords{gamma rays: bursts--- supernovae: general --shock waves}

\maketitle

\section{Introduction}
\label{sec:intro}

In our previous paper \citep{Campana06} the points mentioned in the 
abstract were outlined  only briefly, using order-of-magnitude arguments and 
with very little explanation, due to space limitations.  Here we present a more 
detailed explanation and analysis of the prompt thermal X-ray emission and X-ray 
afterglow, as well as a calculation of the early optical-UV emission, and show 
that some claims made in recent publications \citep{Ghisellini06,Fan06,LiLX06}, 
according to which the observations are inconsistent with the massive wind 
interpretation, are not valid.  

GRB060218 was unique mainly in two respects: it showed a strong thermal X-ray 
emission accompanying the prompt non-thermal emission, and a strong optical-UV
emission at early, $t\lesssim 1$~d, time. We show here that these features, as well as 
the X-ray afterglow, can all be explained by a model in which a radiation mediated
shock propagates through a compact progenitor star into a massive wind. We have 
shown in a separate paper \citep{Wang06} that the prompt non-thermal X-ray emission
can also be explained by this model. 

As detailed below, the prompt thermal X-ray emission can be explained as shock
breakout at a radius of $\sim5\times10^{12}$~cm, which requires a Thomson optical depth
(of the plasma ahead of the shock) of $\tau\approx 1$. Breakout may occur at 
$\sim5\times10^{12}$~cm if this is the stellar progenitor radius. However, since the 
progenitor is likely to be smaller, we suggested the possibility of it being 
surrounded by an optically thick wind. Another possibility is a pre-explosion ejection 
of a small mass, $\sim 10^{-6}M_\odot$, shell.  Here we shall adopt the wind 
interpretation, since it allows one to explain also the X-ray afterglow.  So far, no 
other quantitative physical models have been worked out for the thermal X-ray emission, 
nor for the X-ray afterglow. \citet{Soderberg06} and \citet{Fan06} have suggested that 
the afterglow X-ray emission is due to extended activity of the source, for which there 
is no model or explanation. \citet{Ghisellini06} suggest, for explaining the X-ray 
afterglow, an ansatz consisting of the ad-hoc existence of electrons with some 
prescribed energy distribution, which is different at different times to account for 
the observations, without a model for the dynamics of the plasma or for the electron 
energy distribution.

We note that the radio afterglow of GRB060218 discussed by \citet{Soderberg06} is 
difficult to explain within the context of the current model. Indeed, as pointed 
out by \citet{Soderberg06} and by \citet{Fan06}, it is difficult to explain
the radio afterglow and the X-ray afterglow of GRB060218 as due to emission
from a single shock wave. These authors have thus chosen to construct models
that account for the radio emission only, attributing the X-ray afterglow to
a continued activity of the source of an unexplained nature, and not accounting for
the prompt X-ray emission and for the early optical-UV emission. We adopt a different
approach, showing that the prompt (thermal and non-thermal) X-ray emission, the 
early optical-UV emission, and the late X-ray afterglow can all be explained within
the context of the same model. We argue that it is the radio afterglow, rather than 
all the other components, which remains unexplained, and which should be attributed 
to a different component. Since the radio emission carries only a negligible fraction 
of the energy, and given the large anisotropy of the explosion, it is not difficult 
to imagine the existence of such an additional low energy component.

In \S~\ref{sec:Model} we discuss the observations and our model. In \S~\ref{sec:compare} 
we compare our analysis with those of other authors. Our results and conclusions are
discussed in \S~\ref{sec:discussion}.

\section{Observations and Model}
\label{sec:Model}

\subsection{Thermal X-ray Emission: Shock Breakout}
\label{sec:breakout}

\begin{figure}[htbp]
\includegraphics[width=6.5cm,angle=270]{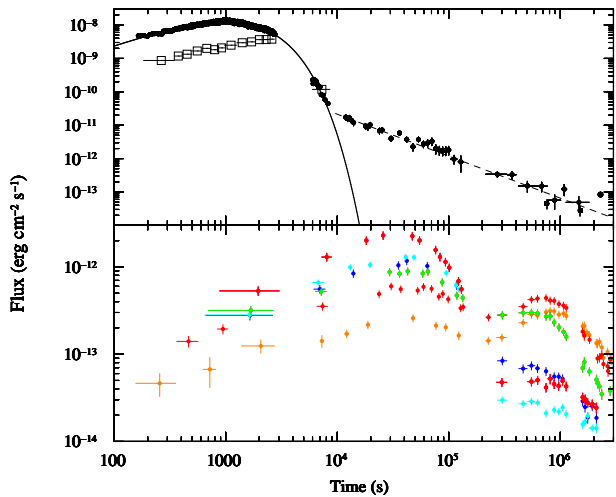}
\caption{{\it Upper panel}: the X-ray light curve (0.3--10 keV). Open black
circles mark Swift-XRT data, filled squares Chandra data (Soderberg et al. 2006) and
filled circle XMM-Newton data (de Luca \& Marshall 2006). Open squares denote the
black body component.
A smooth-burst with exponential decay fits the prompt part (solid line). A power law decay 
with index -1.25 well describes the afterglow tail (dashed line).
Count rate-to-flux conversion factors were derived from
time-dependent spectral analysis. The X--ray light curve has a long, slow power-law rise followed by an
exponential (or steep power-law) decay. At about 10,000 s the light curve breaks to
a shallower power-law decay with index $-1.2\pm0.1$.
{\it Lower panel}: the UVOT light curve. Filled circles of different colors (on-line version)
represent different UVOT filters (from bottom to top between $104-105$ s): $V$ (centered at 544 nm);
$B$ (439 nm), $U$ (345 nm), UVW1 (251 nm); UVM1
(217 nm) and UVW2 (188 nm). Specific fluxes have been multiplied by
their FWHM widths (75, 98, 88, 70, 51 and 76 nm, respectively). }
\label{fig:flux}
\end{figure}

Possibly the most distinguishing feature of GRB060218 is the strong thermal X-ray emission
accompanying the prompt non-thermal emission. The temperature of the thermal
X-ray photons observed up to $3\times10^3$~s is $T\approx0.17$~keV 
\citep[][]{Campana06}.
The integrated flux of the black-body X-ray component, $6\times10^{-6}{\rm erg/cm^2}$
(see figure \ref{fig:flux}), corresponds (using $d=145$~Mpc and after correcting for 
the flux that falls outside the XRT band)
to a thermal X-ray energy of $E_{\rm th}\approx2\times 10^{49}$~erg. This is only an approximate
estimate, and the total thermal energy may be somewhat 
larger, due to the gap in XRT observations
between $3\times10^3$~s and $6\times10^3$~s. 
We consider here a model where this radiation is due to a ``shock breakout".

Supernova (SN) shock waves become radiation-mediated when propagating through the stellar envelope
\citep[see][for review]{WW86}. The energy density behind the shock is dominated by 
radiation, and the mechanism that converts kinetic energy to thermal energy at the 
shock transition is Compton scattering. The optical depth of the shock transition layer 
is $\tau_s\simeq c/v_s$, where $v_s$ is the shock velocity. This thickness is determined 
by the requirement that the time it takes a fluid element to flow through the transition 
layer, $\sim \Delta_s/v_s$, should be comparable to the diffusion time of photons across 
this layer, $\simeq\tau_s\Delta_s/c$.  As the shock propagates outward, the (Thomson) 
optical depth of the plasma lying ahead of the shock decreases, and when this optical 
depth becomes comparable to $\tau_s$, Compton scattering can no longer maintain the shock. 
At this point, radiation escapes ahead of the shock, producing the shock breakout flash 
\citep{Colgate74,Klein78,Ensman92,MM99}, and the shock becomes collisional 
\citep[or collisionless, see][]{WL01}.

Hereafter we use the term "shock breakout" to denote the event of the transition from 
radiation to collisional (collisionless) shock mediation, accompanied by the emission
of radiation. If the optical depth of the wind surrounding the progenitor star is 
small, shock breakout will take place as the shock approaches the stellar surface. It
is for this reason customary to identify shock breakout with the emergence of the shock
from the stellar surface. However, if the optical depth of the wind is large, $>\tau_s$,
breakout would occur once the shock reaches a radius where the wind optical depth drops
to $\tau_s$. 

As argued in \citet{Campana06}, in order to obtain a breakout flash with energy of
$E_{\rm th}\simeq 10^{49}$~erg and temperature $T\approx0.17$~keV, breakout must occur at a radius
of $r\sim5\times10^{12}$~cm, and the shock must be mildly relativistic (which implies
$\tau_s\simeq1$). Since the progenitor star is presumably smaller than this (e.g. if
it is a Wolf-Rayet star), a value of $\tau\simeq1$ at $r\sim5\times10^{12}$~cm may be obtained 
either by assuming that the progenitor is surrounded by a dense wind, or that an outer 
shell of the star was ejected to this radius prior to the GRB explosion. The mass of the 
shell required to obtain $\tau_s\simeq1$ is only $\sim 4\pi r^2/\kappa\sim 10^{-6}
M_\odot$, where $\kappa=0.2{\rm g/cm^2}$ is the Thomson opacity for ionized He. For 
the calculation below we adopt a density profile $\rho\propto r^{-2}$, as would be 
expected for the wind model. The results obtained for the breakout radius, shock
velocity and plasma density at this radius are not sensitive, however, to the details
of the density profile shape.

We consider therefore a shock wave of velocity $\beta c$ driven into the wind 
surrounding the progenitor, whose 4-velocity is $u\equiv\gamma\beta$ where 
$\gamma=(1-\beta^2)^{-1/2}$ is the Lorentz factor. Let us first derive the energy
and temperature of the post-shock radiation. The post-shock temperature $T_d$ is
related to the post-shock pressure by $aT_d^4=3p$, and the observed temperature
is given by $\gamma_d T_d$, where $\gamma_d=(1-\beta_d^2)^{-1/2}$ and $\beta_d c$
is the post-shock plasma velocity ($\beta_d<\beta$). For a strong radiation dominated
shock $p=f u^2\rho c^2$, where $\rho$ is the pre-shock density and $f(u)$ is a factor
of order unity. As we show below, the shock is required to be mildly 
relativistic, $u=$~a few, for which we may approximate $f=0.8$ (for $u\gg1$,
$f=2/3$), and $(u_d/u)^2\approx0.6$ ($(u_d/u)^2=1/2$ for $u\gg1$). 
This defines the observed temperature, $T$, to be
\begin{equation}\label{eq:T}
    T\simeq1 u^{3/2}(a^{-1}\rho c^2)^{1/4}
\end{equation}
(for $u\gg1$, the numerical coefficient is $1/2^{1/4}$ instead of 1).
The energy carried by the radiation
may be estimated by noting that the energy density of the radiation is given, in
the observer frame, by $-p+\gamma_d^2(aT_d^4+p)=(4u_d^2+3)p$, and that the
thickness of the shocked plasma shell may be approximated as $d=(\beta-\beta_d)r$,
where $r$ is the shock radius. This yields
\begin{equation}\label{eq:E}
    \frac{E_{\rm th}}{4\pi r^3\rho c^2}=(\beta-\beta_d)(4u_d^2+3)\frac{p}{\rho c^2}\simeq 0.5 u^2
\end{equation}
(for $u\gg1$, the numerical coefficient is $2/3$ instead of 0.5).

For a wind density profile, the optical depth of the wind at radius $r$ is simply
$\tau(r)=\kappa\rho(r)r$. Expressing the density in terms of $\tau(r)$ and using 
eqs.~(\ref{eq:T}) and~(\ref{eq:E}) we find that the shock velocity at breakout is
\begin{eqnarray}\label{eq:u}
u^2&=&\left(\frac{aT^4}{\sqrt{2\pi}c^3}\right)^{1/7}\left(\frac{\kappa}{\tau}\right)^{3/14}
E_{\rm th}^{1/14}
\nonumber\\
  &=&2.5\left(\frac{T}{0.17\rm keV}\right)^{4/7}E_{\rm th,49}^{1/14},
\end{eqnarray}
and that breakout occurs at a radius
\begin{eqnarray}\label{eq:r}
    R_{ph}&=&\left[\frac{E_{\rm th}^3}{(2\pi)^3aT^4c^4}\right]^{1/7}\left(\frac{\kappa}{\tau}\right)^{2/7}
\nonumber\\
    &=&7.8\times10^{12}\left(\frac{T}{0.17\rm keV}\right)^{-4/7}E_{\rm th,49}^{3/7}\,{\rm cm},
\end{eqnarray}
where $E_{\rm th}=10^{49}E_{\rm th,49}$~erg and we have used $\kappa=0.2{\rm g/cm^2}$ and $\tau=1$ in
the numerical evaluations. Note that the results depends only weakly on the exact values
of $\kappa$ and $\tau$.

Eqs.~(\ref{eq:u}) and ~(\ref{eq:r}) imply that the breakout of the shock takes place 
at a radius $\simeq7.8\times10^{12}$~cm, and that the shock is mildly relativistic 
at breakout, $u^2=2.5$ i.e. $\beta=0.85$. These results are in agreement with the order 
of magnitude estimates given in \cite{Campana06}. Since the shock is found to be mildly 
relativistic, using $\tau=1$ at shock breakout is justified. In order to produce the 
shock that is driven into the wind, the explosion is thus required to produce a mildly 
relativistic shell, $\beta=0.8$, with an energy of a few~$\times10^{49}$~erg.
This is remarkably similar to the case of GRB980425/SN1998bw, for which the ejection of a 
shell of energy $10^{49.7}$~erg and velocity $\beta=0.8$ was inferred from X-ray \citep{W04b}
and radio \citep{Kulkarni98,WL99,CL99,W04a} observations. 

The wind density at the breakout radius, $\rho=1/\kappa r=10^{-12}{\rm g/cm^3}$, is also 
in agreement with our results in \cite{Campana06}. It is straightforward to verify that 
the energy density is dominated by radiation, $aT^4/nT\sim10^9$. For a wind velocity of 
$\sim10^3{\rm km/s}$ it corresponds to a mass loss rate of few~$\times10^{-4}M_\odot/
{\rm yr}$. The relevant mass loss here is that occurring within a day or
less of the explosion. The data currently available on wind mass losses suggesting
${\dot M}\lesssim \hbox{few}\times 10^{-4}M_\odot/{\rm yr}$ refer to much longer 
timescales, for stars well before any explosion \citep{Maeder07}. Physically,
it is however quite plausible that the mass loss increases considerably as the 
evolution of the core rapidly approaches the final collapse, accompanied by a
rapid increase in the luminosity and the envelope expansion rate.

Note that if both the shell and the wind were spherically symmetric, the 
characteristic timescale would be $R_{ph}/c=260$~s (for the inferred post shock velocity, $\gamma_d\beta_d=1.2$, the effects of relativistic beaming are not significant), while the observed timescale of 
the thermal X-ray emission is $\sim 10^3$~s  \citep{Campana06}. However, an
anisotropic shell ejection is a natural expectation in a core collapse GRB, 
since strong rotation is a requisite to make the jet (e.g. \citet{MacFadyen99}).
Even ``normal" core collapse SN simulations show strong rotation-related anisotropy 
in the expanding gas (e.g. \citet{Burrows07}). Thus, the semi-relativistic outer
shell ejected is likely to be anisotropic, either due to an anisotropic explosion, 
or due to being driven by a jet. This, as well as an anisotropic wind profile caused
by rotation \citep{Maeder07} should lead to significant departures from sphericity
in the shock propagation. Anisotropy is in fact a prediction of this model, which is 
supported by the detection of linear polarization \citep{Gorosabel06}. In an anisotropic
shock, however, the timescale is no longer the naive spherical $r/c$ value, but
is rather given by the sideways pattern expansion timescale, which depends on the
angular velocity profile of the anisotropic shell (e.g. at larger angles the shock 
emerges later due a decreasing velocity profile or due to an increasing wind density
away from the symmetry axis, etc.).

\subsection{X-ray afterglow: Wind-shell interaction}
\label{sec:wind-shell}

If the wind shock breakout interpretation is adopted, the subsequent interaction of 
the ejected relativistic shell with the wind is expected to produce an X-ray afterglow
\citep{W04b}. The electrons accelerated to high energy in the collisionless shock driven
into the wind emit X-ray synchrotron radiation, and may also inverse-Compton scatter 
optical-UV SN photons to the X-ray band. 

As the shock wave driven into the wind expands, it heats an increasing amount of mass 
to a high temperature and the initial energy of the ejected shell is transferred to 
the shocked wind. The shell begins to decelerate beyond a radius at which the shocked 
wind energy becomes comparable to the initial shell energy. This occurs at a time 
\citep{W04b}
\begin{equation}\label{eq:t_dec}
    t_{dec}=0.06\frac{E_{k,49}}{\beta^3(\dot{m}/10)}\,{\rm day}.
\end{equation}
Here, $E_k=10^{49}E_{k,49}$~erg is the kinetic energy of the shell and
\begin{equation}\label{eq:mdot}
    \dot{m}\equiv\frac{\dot{M}/v_w}{10^{-5}(M_\odot/{\rm yr})/10^3({\rm km/s})}
\end{equation}
For the wind density and shell velocity inferred in \S~\ref{sec:breakout},
$\beta=0.85$ and $\dot{m}\gtrsim10$, $t_{dec}\lesssim10^4$~s. At $t>t_{dec}$ the energy
is carried by shocked wind plasma, and is continuously transferred to a larger mass of
newly shocked parts of the wind. Thus, the rate at which energy is transferred to 
accelerated electrons is $\approx \epsilon_e E_k/t$, where $\epsilon_{e}$ is the fraction 
of the post-shock thermal energy carried by electrons. If the electrons cool on a time 
scale shorter than the expansion time, $\sim t$, this would lead to a bolometric 
luminosity 
$L_B\approx10^{43}\epsilon_{e,-1}E_{k,49}(t/{1\rm day})^{-1}\,{\rm erg/cm^2/s}$. 
The fraction of this luminosity emitted as X-rays depends on the electron energy 
distribution. For electrons accelerated to a power-law distribution in energy, 
$d\log n_e/d\log\gamma_e\approx-2$, 
$\nu L_\nu\approx L_B/2\log(\gamma_{max}/\gamma_{min})\approx 
10^{42}\epsilon_{e,-1}E_{k,49}(t/{1\rm day})^{-1}\,{\rm erg/cm^2/s}$, where $\gamma_{max}$
and $\gamma_{min}$ are the maximum and minimum electron Lorentz factor. The more 
detailed analysis given in \citet{W04b} of the synchrotron emission of shock 
accelerated electrons gives \citep[][eqs.~(6) and~(7)]{W04b}
\begin{equation}\label{eq:synch}
    \nu L_{\nu,\rm synch}
    \approx3\times10^{41}\epsilon_{e,-1} E_{k,49}t_d^{-1}\,{\rm erg/s}
\end{equation}
for 
\begin{equation}\label{eq:nu_c}
    \nu>\nu_c\approx3\times10^{10}\epsilon_{B,-1}^{-3/2}(\dot{m}/10)^{-3/2}t_d\,{\rm Hz}.
\end{equation}
Here, $t=1t_d$~day, $\epsilon_{e}=10^{-1}\epsilon_{e,-1}$ and $\epsilon_{B}=10^{-1}\epsilon_{B,-1}$
are the fractions of the post-shock thermal energy carried by electrons and magnetic 
field and $\nu_c$ is the cooling frequency, the frequency of synchrotron photons 
emitted by electrons for which the synchrotron cooling time equals $t$ (higher energy 
electrons cool faster and emit higher energy photons). Eq.~(\ref{eq:synch})
implies an X-ray luminosity in Swift's XRT band, 0.3--10~keV, of
\begin{equation}\label{eq:LX}
    L_{X,\rm synch}
    \approx10^{42}\epsilon_{e,-1} E_{k,49}t_d^{-1}\,{\rm erg/s}.
\end{equation}

Let us consider next the inverse-Compton scattering of SN optical-UV photons by the 
shock-heated electrons. The inverse-Compton luminosity emitted by electrons with 
Lorentz factor $\gamma_e$ is given by $L_{IC}\approx\gamma_e^2\tau(\gamma_e)L_{SN}$, 
where $\tau(\gamma_e)$ is the Thomson optical depth of these electrons and $L_{SN}$ 
is the SN luminosity. The lowest energy IC photons are produced 
by the lowest energy, thermal, electrons, 
$(h\nu)_{IC,T}\approx(4\gamma_T^2/3)2$~eV, where $\gamma_T$ is the Lorentz factor of 
the lowest energy electrons. $\gamma_T$ is determined from the post-shock
thermal energy density, $[2/(\Gamma^2-1)]\rho v_s^2\approx\rho v_s^2$ 
where $v_s$ is the shock velocity and $\rho=\dot{M}/4\pi r^2v_w$ ($\Gamma\approx5/3$
is the adiabatic index), through 
$\Lambda n_e\gamma_T m_e c^2=\epsilon_e\rho v_s^2$. Here, $\Lambda$ is a dimensionless
constant, the value of which depends on the exact form of the electron energy 
distribution. For a power-law distribution,
$d\log n_e/d\log\gamma_e\approx-2$, $\Lambda\approx\log(\gamma_{max}/\gamma_{min})$.
 The shock velocity may be inferred from the shock radius, which is 
given by \citep[][eq. 2]{W04b}
\begin{eqnarray}\label{eq:Rs}
    R_s&=&0.73\left(\frac{E_k}{\dot{M}/4\pi v_w}t^{2}\right)^{1/3}
\nonumber\\
    &=&1.8\times10^{15}\left(\frac{E_{k,49}}{\dot{m}/10}\right)^{1/3}
    t_d^{2/3}\,{\rm cm}.
\end{eqnarray}
Using $v_s=(2/3)R_s/t$ we find
\begin{equation}\label{eq:gamma_m}
    \gamma_T\approx80\epsilon_{e,-1}\Lambda^{-1}
    \left(\frac{E_{k,49}}{\dot{m}/10}\right)^{2/3}t_d^{-2/3},
\end{equation}
and
\begin{equation}\label{eq:nu_IC}
    (h\nu)_{IC,T}\approx 15 \epsilon_{e,-1}^2\Lambda^{-2}
    \left(\frac{E_{k,49}}{\dot{m}/10}\right)^{4/3}
    t_d^{-4/3}\, {\rm keV}.
\end{equation}
Thus, on a time scale of a days IC scattering by the lowest 
energy (thermal) electrons is expected to contribute to the X-ray flux. It is 
important to note here that the electrons producing the X-ray synchrotron flux, 
eq.~(\ref{eq:LX}), are of much higher energy, $\gamma_e\gg\gamma_T$, and lie at 
the high energy part of the accelerated electron energy distribution. 

The Thomson optical depth of the thermal electrons is approximately given by
$\kappa M(r)/4\pi r^2=\kappa\rho r$, where $M(r)$ is the wind mass accumulated up to 
$r$.  Assuming that the thermal electrons do not lose all their energy by IC scattering
on a time scale shorter than the expansion time $t$, the IC luminosity produced by the 
thermal electrons is 
\begin{eqnarray}\label{eq:L_IC}
    L_{IC,T}&\approx& \kappa\rho r \gamma_T^2 L_{SN}
\nonumber\\
    &\approx&9\times10^{42} \frac{\epsilon_{e,-1}^2}{\Lambda^{2}}
   {E_{k,49}}\frac{L_{SN}}{10^{42.5}{\rm erg/s}}
    t_d^{-2/3}\,{\rm erg/s}.
\end{eqnarray}
The IC luminosity given by eq.~(\ref{eq:L_IC}) may exceed, depending on the values
of $\Lambda$ and $\epsilon_e$, the luminosity given by eq.~(\ref{eq:LX}). 
$L_{IC,T}>L_{X,\rm synch}$ is not, of course, a valid result, 
since $L_{X,\rm synch}$ given in eq.~(\ref{eq:LX}) is the luminosity 
obtained assuming that the electrons lose to radiation all the energy they gained 
from the shock.  The IC luminosity is thus limited by $L_{X,\rm synch}$, and 
$L_{IC,T}>L_{X,\rm synch}$ simply implies that the thermal electrons lose all 
their energy to IC scattering on a short time scale. 

Let us compare the predicted X-ray afterglow with the observed one.
The observed X-ray afterglow, following the prompt emission which ends at 
$\sim 10^4$~s, is well approximated by (see figure~\ref{fig:flux})
$f_X=10^{-12}t_d^{-1.2\pm0.1}{\rm erg/cm^2/s}$, which corresponds to
$L_X=2\times10^{42}t_d^{-1.2\pm0.1}{\rm erg/s}$. 
This is in excellent agreement 
with the predictions of eqs.~(\ref{eq:LX}) and ~(\ref{eq:L_IC}) for 
$E_k=$~few~$\times10^{49}$~erg. The fact that the energy of the shock 
driven into the wind inferred from the X-ray afterglow, $E_k$, is comparable to that
of the thermal X-rays, $E_{\rm th}$, supports our model, in which both are due
to the same shock driven into the wind.

At early time, 
$t<1$~day, the emission is expected to be dominated by the synchrotron component, and
for the power-law energy distribution of electrons, $d\log n_e/d\log\gamma_e\approx-2$, 
the X-ray spectrum is expected to follow $\nu L_\nu\propto\nu^{0}$. On a time
scale of a few days, the emission is expected to be dominated by IC scattering of 
thermal SN photons by thermal shock electrons. At this stage, the X-ray spectrum
may become steeper, reflecting the energy distribution of the lowest energy electrons 
heated by the shock. These results are consistent with the $\nu L_\nu\propto\nu^{0}$ 
X-ray spectrum measured at $t\sim1$~day, and with the indication, based on
an XMM-Newton observation, that at a later time, $t\sim3$~day, the spectrum is steeper, 
$\nu L_\nu\propto\nu^{-1.3\pm0.6}$ \citep{deLuca06}. 

The following point should be made here. As the shock speed $\dot{R_s}$ decreases with time, shells ejected with velocities lower than that of the fast, $\beta\simeq0.8$, shell may "catch up" with the shock ($R_s/ct=0.6 t_d^{-1/3}$ for $E_{k,49}=1$ and $\dot{m}=10$). This may lead to increase with time of the shock kinetic energy $E_k$, and thus to a modification of the X-ray light curve. The fact that the X-ray luminosity decreases roughly as $1/t$ up to $\sim10$~d implies that the energy in shells ejected with $\beta\gtrsim0.3$ is not much larger than that of the fast shell, i.e. not much larger than $10^{49}$~erg.

\subsection{Early optical-UV emission: Envelope cooling}
\label{sec:UV}

As the SN shock propagates through the stellar envelope, it heats it to $\sim1$~keV. 
As the envelope expands, the photosphere propagates into the envelope, and we see 
deeper shells with lower temperatures. We derive here a simple analytic model for 
the photospheric radius and temperature, based on which we can derive approximately
the flux and temperature of the escaping radiation. We assume that the density of 
the stellar envelope near the stellar surface is given by 
\begin{equation}
\label{eq:rho0}
\rho_0(r)=\rho_{1/2}\delta^n,
\end{equation}
where $\delta\equiv(1-r/R_*)$ ($R_*$ is the stellar radius), and that the the plasma 
ahead of the photosphere is highly ionized He, so that $\kappa=0.17\,{\rm cm^2/g}$. 
The velocity of the shock as it propagates through the envelope is approximately 
given by \citep{MM99,TMM01}
\begin{eqnarray}
\label{eq:vs}
  v_s&=&0.8\left(\frac{E_{ej}}{M_{ej}}\right)^{1/2}\left(\frac{M_{ej}}{\rho r^3}\right)^{0.2}
\nonumber\\
     &\xrightarrow[r\rightarrow R_*]{}&0.8\left(\frac{E_{ej}}{M_{ej}}\right)^{1/2}
              \left(\frac{M_{ej}}{\rho_{1/2} R_*^3}\right)^{0.2}\delta^{-0.2n},
\end{eqnarray}
where $M_{ej}$ is the mass of the ejected envelope, and $E_{ej}$ is the energy deposited in the 
envelope. Since the shock is radiation dominated, the energy density behind the shock 
is given by $u_{\rm rad,0}/3=(6/7)\rho_0 v_s^2$,
\begin{equation}
\label{eq:T0}
 aT_0^4(\delta)=u_{\rm rad,0}(\delta)=\frac{18}{7}\rho_0(\delta) v_s(\delta)^2.
\end{equation}

After the shock breaks through the envelope, the envelope begins to expand and cool. 
It is useful do label the shells with Lagrangian coordinates, defining
$\delta_m(\delta)M_{ej}$ as the (time independent) mass that lies ahead of a shell 
originally located at $r=(1-\delta)R_*$. \citet{MM99} show that the final velocity 
(after acceleration due to the adiabatic expansion) of each shocked shell, $v_f$, 
is related to the shock velocity approximately by $v_f(\delta_m)=2v_s 
[\delta(\delta_m)]$. After significant expansion, $v_ft\gg R_*$, the radius of each 
shell is given by $r(\delta_m,t)=v_f(\delta_m)t$. Given $r(\delta_m,t)$ it is 
straightforward to derive the time dependent shell density. The shell's (time 
dependent) temperature is then determined by $T/T_0=(\rho/\rho_0)^{4/3}$, which 
holds for adiabatic expansion. After some tedious algebra we find that, for 
$v_ft\gg R_*$, the photosphere is located at
\begin{equation}
\label{eq:photo}
 \delta_{m,\rm ph}(t)=3.8\times10^{-3}f_\rho^{-0.07}\frac{E_{ej,51}^{0.8}}{(M_{ej}/M_\odot)^{1.6}}t_d^{1.6}.
\end{equation}
Here, $E_{ej}=10^{51}E_{ej,51}$~erg and $f_\rho\equiv \rho_{1/2}/\bar{\rho_0}$ 
is the ratio between $\rho_{1/2}$ and the average envelope density $\bar{\rho_0}$. 
Although $f_\rho$ depends on the structure of the progenitor star far from the 
surface, where eq.~(\ref{eq:rho0}) no longer holds, the results are very insensitive 
to its value. Using eqs.~(\ref{eq:vs}) we find that the radius of the photosphere is
\begin{equation}
\label{eq:r_photo}
 r_{\rm ph}(t)=3.2\times10^{14}f_\rho^{-0.04}\frac{E_{ej,51}^{0.4}}{(M_{ej}/M_\odot)^{0.3}}t_d^{0.8}\,{\rm cm},
\end{equation}
and using eq.~(\ref{eq:T0}) we find that the temperature of the photosphere is
\begin{equation}
\label{eq:T_photo}
 T_{\rm ph}(t)=2.2f_\rho^{-0.02}\frac{E_{ej,51}^{0.02}}{(M_{ej}/M_\odot)^{0.03}}R_{*,12}^{1/4}
t_d^{-0.5}\,{\rm eV}.
\end{equation}
Here, $R_*=10^{12}R_{*,12}$~cm.

\citet{Campana06} find $R\approx3\times10^{14}$~cm and $T\approx3$~eV at $t=10^5$~s. 
This is clearly consistent with the cooling envelope interpretation. We note that at 
$t=10^5$~s the emission originates from a shell of mass $\simeq10^{-3}M_\odot$, in 
excellent agreement with our rough estimate in \citep{Campana06}.  At the time 
$t\ll10^5$~s optical photons are in the Rayleigh-Jeans tail, and the flux is 
approximately given by 
$f_\nu\approx2\pi(\nu/c)^2T_{\rm ph}(r_{\rm ph}/D)^2$, where $D$ is the distance to 
GRB060218. Using eqs.~(\ref{eq:r_photo}) and~(\ref{eq:T_photo}) we find
\begin{eqnarray}
\label{eq:LRJ}
  \nu f_\nu\approx 1.3\times10^{-13} f_\rho^{-0.1} \frac{E_{ej,51}^{0.8}}{(M_{ej}/M_\odot)^{0.6}}R_{*,12}^{1/4}
\nonumber\\
\times\left(\frac{\nu}{5\times10^{14}{\rm Hz}}\right)^3t_4^{1.1}\,{\rm erg/cm^2/s},
\end{eqnarray}
where $t=10^4t_4$~s.
This is consistent with the optical flux observed at early time, 
see figure~\ref{fig:flux}
(note that our 
derivation holds only for $r_{\rm ph}\gg R_*$, i.e. for $t>10^3$~s). 

A more detailed study of the 
optical-UV early light could provide interesting constraints on the size and 
density distribution near the surface of the progenitor star. Such a detailed 
analysis would require taking into account effects neglected here (e.g., photon 
diffusion, which may be important on $>1$~day time scale, anisotropy, etc.) 
and is beyond the scope of the current manuscript.

Note that we have neglected the effects of photon diffusion in the above derivation. We do not expect diffusion to play an important role, due to the following argument. The size of a region around $r(\delta_m,t)$ over which diffusion has a significant effect is $\Delta(\delta_m,t)=\sqrt{ct/3\kappa\rho(\delta_m,t)}$. Thus, the value of $\delta_m$ up to which diffusion affects the radiation field significantly, $\delta_{D}$, is determined by $\Delta(\delta_m=\delta_{D},t)/r(\delta_m=\delta_{D},t)=1$, which gives 
\begin{equation}\label{eq:rRatio}
    \frac{r(\delta_m=\delta_{D},t)}{r_{\rm ph.}}=1.2 f_\rho^{-0.005}\frac{E_{ej,51}^{0.05}}{(M_{ej}/M_\odot)^{0.03}}t_d^{-0.03}.
\end{equation}
This implies that photon diffusion is not expected to significantly modify the predicted light curve
\citep[Applying, e.g., the approximate self-similar diffusion wave solutions of][to our density and velocity profiles yields a luminosity that differs by $<25\%$ from that derived neglecting diffusion]{Chevalier1992}.

\subsection{Radio emission}
\label{sec:radio}

For the massive wind discussed in our model, the synchrotron self-absorption optical
depth is very large at radio frequencies. The characteristic frequency of  synchrotron
photons emitted by the lowest energy, $\gamma_e\sim\gamma_T$, electrons is
$\nu_m\approx\gamma_T^2 eB/2\pi m_ec$, where the magnetic field is given by
$B^2/8\pi=\epsilon_B\rho v_s^2$. Using eqs.~(\ref{eq:Rs}) and~(\ref{eq:gamma_m})
we have
\begin{equation}\label{eq:nu_m}
    \nu_m\approx5\times10^{11}\Lambda^{-2}\epsilon_{B,-1}^{1/2}\epsilon_{e,-1}^{2}
    E_{k,49}^{4/3}(\dot{m}/10)^{-5/6}t_d^{-7/3}\,{\rm Hz}.
\end{equation}
On a timescale of a few days we expect therefore the radio frequency,
$\nu\sim10$~GHz, to be in the range $\nu_m\lesssim\nu<\nu_c$ (see eq.({\ref{eq:nu_c})),
i.e. we expect the Lorentz factor of electrons dominating the emission and absorption
of radio waves to be in the range $\gamma_T\lesssim\gamma_\nu<\gamma_c$ (where $\gamma_c$
is the Lorentz factor of electrons with cooling time comparable to the expansion time). 
In this case, the synchrotron self-absorption optical depth is given by
$\tau_\nu=[e^3B/2\gamma_\nu(m_ec\nu)^2]n_e(\gamma_\nu)\Delta$, 
where $n_e(\gamma_\nu)$ is the number 
density of electrons with Lorentz factor $\gamma_\nu$ and $\Delta$ is the thickness
of the shocked wind shell. For a power-law distribution, $dn_e/d\gamma_e\propto\gamma_e^{-2}$,
we have $n_e(\gamma_\nu)\approx n_e\gamma_T/\gamma_\nu$, and using
$n_e\Delta=\rho r/2m_p$ for the electron column density we have
\begin{equation}\label{eq:tau}
    \tau_\nu\approx10^{7}\frac{\epsilon_{B,-1}\epsilon_{e,-1}}{\Lambda}
    E_{k,49}^{1/3}(\dot{m}/10)^{5/3}\left(\frac{\nu}{10\,{\rm GHz}}\right)^{-3}t_d^{-10/3}.
\end{equation}
This large optical depth leads to a strong suppression of the radio synchrotron flux,
compared to the X-ray synchrotron flux given in eq.~(\ref{eq:synch}),
\begin{eqnarray}\label{eq:radio}
    \frac{(\nu L_\nu)_{\rm radio}}{(\nu L_\nu)_X}\approx
    &10^{-4}&\frac{\Lambda}{\epsilon_{B,-1}^{1/4}\epsilon_{e,-1}}
    E_{k,49}^{-1/3}(\dot{m}/10)^{-11/12}
    \nonumber\\
    &\times&\left(\frac{\nu}{10\,{\rm GHz}}\right)^{7/2}(t_d/10)^{17/6}.
\end{eqnarray}

The flux ratio measured at few to 20~days is 
$(\nu L_\nu)_{\rm radio}/(\nu L_\nu)_X\approx10^{-2.5}$ \citep{Soderberg06}.
Thus, as noted in the introduction, the observed radio flux is higher than predicted
by this model, and should be explained as a different component (see also
\S \ref{sec:fan06}).

\section{Comparison with other authors}
\label{sec:compare}

\subsection{Ghisellini et al. 2006}
\label{sec:ghis06}

\citet{Ghisellini06} argue that the shock breakout interpretation is not valid,
since the optical-UV emission at few~$\times10^3$~s is higher than the extrapolation 
to low frequencies of the Planck spectrum which fits the thermal X-ray emission at 
the same time. As we have pointed out in \citep{Campana06}, and as explained in 
detail in \S~\ref{sec:Model}, the thermal X-ray emission and the early optical-UV 
emission originate from different regions. The thermal X-rays originate from a 
(compressed) wind shell of mass $\sim10^{-6}M_\odot$, while the optical-UV emission 
originates from the outer shells of the (expanding) star. At $\sim10^4$~s the 
optical-UV radiation is emitted from a shell at a "depth" of $\sim10^{-4}M_\odot$
from the stellar edge. The optical-UV and the thermal X-rays need not correspond 
to the same Planck spectrum. As we have explained in \citep{Campana06}, and in more 
detail in \S~\ref{sec:Model}, if the explosion had been isotropic, the thermal X-ray
emission should have disappeared altogether on a time scale of few hundred seconds,
at which time \citet{Ghisellini06} compare the X-ray and UV emission. The fact that
the thermal X-ray emission is observed over a few thousand seconds can be accounted 
for by assuming an anisotropic explosion. This assumption, made explicitly both in 
\citep{Campana06} and here, is ignored by \citet{Ghisellini06}, whose criticism
is framed in the context of a spherical model, which we never assumed.

Concerning their own model, in their preferred explanation for the properties 
of this burst \cite{Ghisellini06}, unlike in our model, do not include any dynamics 
in their 
scenario, and adopt different ad-hoc parameters (e.g. for the electron 
distribution) at different times to account for the observations.

\subsection{Li 2006}
\label{sec:li06}

\citet{LiLX06} argues that in order to explain the ``temperature and the total 
energy of the blackbody component observed in GRB 060218 by the shock breakout,
the progenitor WR star has to have an unrealistically large core radius ...
larger than $100R_\odot$". The results of \citet{LiLX06} are in fact consistent 
with our analysis. The ``core radius", $R_c$, adopted by this author refers not to
the hydrostatic core radius of the star, but rather to the radius at which the
optical depth equals 20. In non-LTE modeling of the winds of WR-stars, 
$R_c$ is located within the wind, typically near the sonic point. In fact, 
in some cases the wind velocity is already supersonic at $R_c$ (e.g. Hamann \& 
Koesterke 1998). $R_c$ is much larger than the hydrostatic core 
radius $R_{hc}$ obtained in evolutionary calculations (e.g. Schaerer \& Maeder 
1992). As shown by Nugis \& Lamers (2002), 
$R_c/R_{hc}$ varies with the spectral type of WR stars, increasing from about 2 
for early types to about 20 for late types (see table 5 of Nugis \& Lamers 
2002). 

Since Li (2006) considers in his calculations models where the ratio between the
wind photospheric radius $R_{ph}$ and $R_c$ is $1<R_{ph}/R_c<3$ (using his 
notation, for $b$=5 the choice of $\epsilon=10^{-5}$ to $10^{-2}$ 
corresponds to $1<R_{ph}/R_c<3$, see his fig. 3), his statement that
$R_c\sim100R_\odot$ is required implies that a wind with a photospheric 
radius of $7\times10^{12}$~cm to $2\times10^{13}$~cm is needed to account 
for the thermal X-ray emission of GRB 060218 as shock breakout.
This is consistent with our eq.~(\ref{eq:r}). 

It is important to emphasize that the relevant radius is not $R_c$, but rather
the wind photospheric radius, $R_{ph}$, where $\tau=1$. It is the photospheric 
radius which needs to be large in order to allow sufficiently large shock 
breakout energy. $R_{ph}$ is larger than $R_c$ by a factor of 2--10, depending
on the model adopted for the wind velocity profile. Nugis \& Lamers 2002 
present late type models with $R_c/R_\odot>30$, and a large fraction of the stars 
analyzed in Hamann \& Koesterke (1998) have $R_c/R_\odot>25$ (see their table 2). 
$R_{ph}$ larger than $100R_\odot$ may therefore be obtained for stars with large 
mass loss rates. 

It should be emphasized that even if the wind photospheric radius required for 
GRB 060218 had turned out to be larger by a factor of a few than the largest 
$R_{ph}$ of known WR stars, this would not have been a strong argument against 
the wind interpretation of GRB 060218, since GRB progenitors may well be more 
extreme than normal WR stars. Clearly, not all WR stars end their lives as GRBs. 
Moreover, it should be remembered that practically nothing is known about the 
mass loss on a day time scale prior to the explosion, which determines the wind 
density at the relevant radii.

\subsection{Fan et al. 2006}
\label{sec:fan06}

\citet{Fan06} raised different criticisms in different versions of their paper. 
In the first version, astro-ph/0604016v1, they argued that the presence of a massive 
wind would result in strong optical emission at early time, which is inconsistent 
with observations.  While this claim was retracted in subsequent versions of their 
manuscript, it may be worthwhile to clarify this issue here. The synchrotron emission 
given in eq.~(\ref{eq:synch}) holds for frequencies above the cooling frequency, 
$\nu>\nu_c$. Since the cooling frequency is low for a massive wind, see equation
(\ref{eq:nu_c}), we expect in this model a similar X-ray and optical-UV luminosity, 
$\nu f_\nu\approx 3\times10^{-13} (t/{1\rm day})^{-1}\,{\rm erg/cm^2/s}$ (corresponding 
to a flux of $F_{UV}\approx 1\times10^{-13}(t/{1\rm day})^{-1}\,{\rm erg/cm^2/s}$ in 
the $0.2\mu$ band of Swift's UVOT). This predicted flux exceeds the observed flux
at $t<10^4$~s, hence the statement of \citet{Fan06}. However, as pointed out in 
\S~\ref{sec:wind-shell}, eq.~~(\ref{eq:synch}) holds only for times $t>t_{dec}
\sim10^4$~s (and the luminosity is lower at earlier times).
Moreover, at $t<10^4$~s the large X-ray luminosity, $L_X\sim10^{46}{\rm erg/s}$, would
suppress the synchrotron emission of the shock accelerated electrons. As explained in 
\S~\ref{sec:wind-shell}, the cooling time of shock accelerated electrons due to 
synchrotron emission is short compared to the shock expansion time. At early times,
$t<10^4$~s, the ratio of the magnetic energy density, $\epsilon_B E/r^3$, to the 
X-ray energy density, $L_X/4\pi r^2c$, is 
$4\pi\epsilon_B E c/L_X r\sim 10^{-3.5}\epsilon_{B,-1} E_{k,49}(r/10^{15}{\rm cm})$,
which implies that the IC cooling time is much shorter than the synchrotron cooling
time, suppressing the synchrotron emission in the optical band.

Second, \citet{Fan06} argue \citep[see also][]{Soderberg06} that radio observations
rule out the existence of a massive wind. As mentioned in the introduction, indeed 
the radio and the X-ray afterglow are difficult to explain in the framework of a 
single shock. However, the radio observations alone cannot be used to rule out 
the existence of a massive wind, since  radio observations do not allow one to 
determine the explosion parameters. This is illustrated by the fact that estimates 
for the kinetic energy based on modeling the radio data alone range from 
$\sim10^{48}$~erg to $\sim10^{50}$~erg, and that the ambient medium density estimates 
range from $\sim10^0{\rm cm}^{-3}$ to $\sim10^2{\rm cm}^{-3}$ \citep{Fan06,Soderberg06}.

Finally, \citet{Fan06} argue that in the presence of a massive wind, the radio flux
would be higher than observed. As explained in \S~\ref{sec:radio}, our problem is
quite the opposite: our model flux would be too low to account for the observed 
radio emission. 
\citet{Soderberg06} and \citet{Fan06} have chosen to construct 
models of GRB060218 which concentrate on accounting for the radio emission, while
attributing the X-ray afterglow to a continued activity of the source of an 
unexplained nature, and de-emphasizing the importance of prompt X-ray emission and 
the early optical-UV emission.  However, the radio emission represents a negligible 
fraction of the total energy. Here, and in \citet{Campana06}, we have  adopted a 
different approach, which is that the prompt (thermal and non-thermal) X-ray emission, 
the early optical-UV emission, and the late X-ray afterglow can all be explained 
within the context of the same model, which is based on the energetics of the
early phases of the explosion. We argue that it is the radio afterglow, rather than
all the other components, which plays a lesser role, and which may be attributed
to a different component. Given the very low energy of the radio emission and the
the large anisotropy expected in the explosion, such an additional low energy radio 
component is imaginable, whose role is unlikely to be important in determining
the characteristics of the early high energy emission.

\section{Discussion}
\label{sec:discussion}

We have discussed a comprehensive model of the early X-ray and optical/UV behavior 
of the GRB060218/SN2006aj system, which provides the quantitative justification
for the interpretation outlined in \citet{Campana06}, as well as a number of 
additional points. The most exciting features of this event were that it showed 
a strong thermal X-ray component as well as a strong optical/UV component in its 
early phases, at $t\lesssim 1$ day, transitioning later to a more conventional X-ray 
and optical afterglow, and a radio afterglow. We have shown that the early X-ray/O/UV 
behavior can be understood 
in terms of a mildly relativistic radiation-mediated shock which breaks out of a 
(Thomson) optically thick wind produced by the progenitor star, leading to the
observed thermal X-rays (see \S~\ref{sec:breakout}). 
The early optical/UV behavior arises as the shocked stellar envelope expands 
to larger radii (\S~\ref{sec:UV}), 
and the X-ray afterglow arises from synchrotron and inverse-Compton emission 
of electrons accelerated by the propagation of the shock further into the wind 
(\S~\ref{sec:wind-shell}). A detailed analysis of the optical-UV emission may
therefore provide stringent constraints on the progenitor star.

The thermal X-ray emission requires a mildly relativistic shock, $\beta\approx0.8$,
carrying $\gtrsim10^{49}$~erg, driven into a massive wind characterized by a mass 
loss rate of a few~$\times10^{-4}M_\odot/{\rm yr}$ for a wind velocity of $10^3
{\rm km/s}$. The later X-ray afterglow is consistent with emission due to the
propagation of this shock into the wind beyond the breakout radius, where the shock
becomes collisionless. This situation is very similar to that of GRB980425/SN1998bw,
for which the X-ray and radio afterglow are interpreted as emission from a shock 
driven into a (much less massive) wind by the ejection of a shell of energy 
$10^{49.7}$~erg and velocity $\beta=0.8$ \citep[][and references therein]{W04b}. 

The early optical-UV emission is consistent with the expansion of the outer
part, $\delta M\sim10^{-3}M_\odot$, of the stellar envelope, which was heated 
to a high temperature by the radiation dominated shock as it accelerates in its
propagation towards the stellar edge. It is important to note, however, that the 
acceleration of the shock near the stellar surface is not sufficient to account 
for the large energy, $\gtrsim10^{49}$~erg, deposited in the mildly relativistic, 
$\beta\approx0.8$, component, which is required to have a mass of $\sim10^{-5}
M_\odot$. For the parameters inferred from the SN2006aj light curve,
$E_{ej}\simeq2\times10^{51}$~erg and $M_{ej}\simeq2M_\odot$ \citep{Mazzali06}, 
the energy predicted to by carried by $\beta\ge0.8$ is less than
$10^{45}$~erg \citep[see fig. 6 of][]{TMM01}.
This suggests that the
mildly relativistic component is driven not (only) by the spherical SN shock 
propagating through the envelope, but possibly by a more relativistic component 
of the the explosion, e.g. a relativistic jet propagating through the star.

An important factor in our interpretation of the early X-ray emission is the 
anisotropy of this shock, which leads to a characteristic timescale of the 
thermal X-ray emission controlled by the sideways pattern expansion speed,
rather than by a simple radial line of sight velocity. A significant anisotropy 
of the mildly relativistic shock component is expected for various reasons, e.g.
as a result of being driven by a jet, or due to propagation through an anisotropic 
envelope and into an anisotropic wind caused by fast progenitor rotation, etc.
The simplest explanation for the prompt gamma-ray behavior may be that it is due to 
a relativistic jet, which can contribute to the anisotropy of the mildly relativistic 
shock propagating through the stellar envelope and the mildly relativistic ejected 
shell. In this case, the contribution of the relativistic jet to the late 
($t \gtrsim 1$ day) X-ray and optical afterglow is not the dominant effect
\citep[c.f.][]{Liang06}. However, a possible explanation of the prompt non-thermal
emission, both in GRB980425 and GRB060218, may be that it is due to repeated IC 
scattering at breakout, as suggested in \citet{Wang06} \citep[c.f.][]{Dai06}. In this 
case there may be no highly relativistic jet- it may either be mildly relativistic to 
begin with, or it may have been relativistic but it was choked, only its mildly 
relativistic bow shock emerging from the star.}

If our interpretation is correct, the importance of the early thermal X-ray 
component is that it represents the first detection ever of the breakout of a
supernova shock from the effective photosphere of the progenitor star. Since
this is a GRB-related supernova, a strong stellar wind can be expected, which
results in the breakout photospheric radius being in the wind, rather than
in the outer atmosphere of the star. This provides a potentially valuable
tool for investigating the physical conditions, mass loss and composition of
the long GRB progenitors in the last few days and hours of their evolution
prior to their core collapse.

\acknowledgements{We are grateful to our Swift colleagues for stimulating 
collaborations, and to NASA NAG5-13286 (PM), ISF and Minerva (EW),
and ASI I/R/039/04 (SC) for partial support.} We thank the anonymous
referee for useful comments that lead to several improvements of the manuscript.

\end{document}